\documentclass[aps,10pt]{revtex4-1}
\usepackage{amsmath,amsfonts,amssymb,amscd,amsxtra,amsthm}
\usepackage{graphicx}
\usepackage[T1]{fontenc}
\usepackage{bm}
\usepackage{hyperref}
\usepackage{multirow}
\usepackage{booktabs}
\usepackage[normalem]{ulem} 

\renewcommand\sout{\bgroup \color{red} \ULdepth=-.5ex \ULset}

\hypersetup{pdfpagemode={UseOutlines},
  bookmarksopen=true,
  bookmarksopenlevel=0,
  hypertexnames=false,
  colorlinks=true,
  citecolor=blue,
  linkcolor=blue,
  urlcolor=blue,
  allcolors=blue,
  pdfstartview={FitV},
  unicode,
  breaklinks=true,
}

\begin{document}
\preprint{}
\title{Systematic analysis of the nuclear absorption effect 
on the cross section of the knockout reaction}
\author{Sang-In Shim}
\email[E-mail: ]{ssimr426@korea.ac.kr}
\affiliation{Center for Extreme Nuclear Matters (CENuM),
Korea University, Seoul 02841, Republic of Korea}
\affiliation{Research Center for Nuclear Physics (RCNP),
Osaka University, Ibaraki, Osaka, 567-0047, Japan}
\author{Kazuki Yoshida}
\affiliation{Advanced Science Research Center, 
Japan Atomic Energy Agency, Tokai, Ibaraki 319-1195, Japan}
\author{Kazuyuki Ogata}
\affiliation{Department of Physics, Kyushu University, 
Fukuoka 819-0395, Japan}
\affiliation{Research Center for Nuclear Physics (RCNP),
Osaka University, Ibaraki, Osaka, 567-0047, Japan}
\date{\today}
\begin{abstract}
  Recent studies on nucleon and alpha knockout reactions have shown that the distorted-wave impulse
approximation (DWIA) is a simple and accurate method to describe these reactions.
  As it has been argued for decades, the nuclear absorption is one of the most important ingredients of the DWIA calculation.
  In this work, we systematically investigate the absorption effects on the cross sections of the nucleon and alpha knockout reactions. 
  To do this, we calculate the ratio of the cross sections of the DWIA and plane-wave impulse approximation (PWIA) and examine its dependence on the mass number and single-particle orbital of the knocked-out particles. 
  We will discuss the specific characteristics of the absorption effect for each reaction.
\end{abstract}
\maketitle

\section{Introduction}
For more than half a century, the proton-induced nucleon knockout reaction, ($p,p$N), has been a powerful spectroscopic tool for studying the single-particle (s.p.) picture of nuclei~\cite{Jacob66,Jacob73,Kitching85};
see also reviews~\cite{Wakasa17,Aumann21_etal}.
In recent years, the s.p. nature and the shell evolution~\cite{Otsuka01,Otsuka05,Otsuka13,Otsuka16} in neutron-rich nuclei have been intensively studied at RIKEN RIBF~\cite{Santamaria15_etal,Paul17_etal,Flavigny17_etal,Chen17_etal,Lettmann17_etal,Shand17_etal,Olivier17_etal,Cortes18_etal,Liu18_etal,Liu19_etal,Paul19_etal,Taniuchi19_etal,Elekes19_etal,Chen19_etal,Cortes20_etal,Cortes20_2_etal,Sun20_etal,Lokotko20_etal,Lizarazo20_etal,Frotscher20_etal,Juhasz21_etal,Juhasz21_V63_etal,Browne21_etal,Linh21_etal,Koiwai22_etal,Gerst22_etal,Enciu22_etal,Elekes22_etal}.
It should also be mentioned that regarding the quenching of the spectroscopic strength~\cite{Gade08_etal,Tostevin14}, the nucleon knockout reaction of a wide range of carbon, nitrogen, and oxygen isotopes has been measured at the R$^{3}$B-LAND setup at GSI~\cite{Panin16_etal,Atar18_etal,Fernandez18_etal,Holl19_etal} and RIKEN~\cite{Kawase18_etal}.
In Ref.~\cite{Phuc19}, these data were reanalyzed using the distorted-wave impulse approximation (DWIA) framework~\cite{Chant77,Chant83}, and various corrections and uncertainties within the standard DWIA formalism on the ($p,p$N) cross sections were systematically investigated.

In the determination of the s.p. wave function using the knockout reaction cross section, the absorption effect in the distorted waves of the scattering particles plays a crucial role.
Since the absorption makes the reaction peripheral, its effect on the reaction observables depends on the s.p. property of a struck nucleon, e.g., its orbital angular momentum $l$, radial quantum number $n$, nucleon separation energy, and its spatial distribution.
Similarly to ($p,p$N), the proton induced $\alpha$ knockout reaction, ($p,p\alpha$), is a suitable probe for the $\alpha$ clustering in nuclei~\cite{Yoshida19,Taniguchi_48Ti}.
It is shown that the ($p,p\alpha$) reaction is peripheral even in the case of light nuclei, $^{20}$Ne($p,p\alpha$)$^{16}$O~\cite{Yoshida19}, and there is almost no contribution from the internal region in medium and heavy mass cases~\cite{Yoshida16,Yoshida18,Taniguchi_48Ti,Yoshida22}.
This is because the mean-free-path of $\alpha$ in a nucleus is short and the absorption effect is much stronger than that of a nucleon. 

In Sec. 5.5. and Fig. 31 of Ref.~\cite{Wakasa17}, the absorption effect has been discussed using the ratio of the DWIA and PWIA (plane-wave impulse approximation) s.p. cross sections. However, the previous work included only a small number of cases. Therefore, a systematic analysis of the absorption effect and its s.p. orbital dependence is required. 
For this purpose, in this paper, we investigate the ratio of the DWIA and PWIA cross sections (the DW-PW ratio) of the nucleon and the $\alpha$ knockout reactions in more diverse atomic nuclei, particularly focusing on the s.p. and cluster orbital dependence.
It can be shown that the PWIA cross section roughly corresponds to the total elastic cross section of the elementary process~\cite{Ogata:2015eta, Mario_private}. 
Systematic knockout reaction experiments are planned for the future, and the DW-PW ratios in this study will be helpful in estimating the cross sections of the knockout reaction from those of the elementary process prior to the actual experiments.

The content of this paper is as follows. 
In section II, we briefly explain the framework of DWIA, PWIA, and the cross sections derived from those methods. 
The numerical inputs for the ($p$,$p$N) and ($p$,$p\alpha$) reactions and information about the target nuclei and the s.p. orbitals are also given there.
In section III, we present the numerical results of the DW-PW ratio for the ($p$,$p$N) and ($p$,$p\alpha$) reactions. 
We discuss the target, s.p. orbital, and incident energy dependences of the DW-PW ratio. 
The final section is devoted to summary and conclusion.

\section{DWIA framework}
\subsection{Knockout cross section}
The factorized form of the DWIA framework is used for the present calculations.
Since we follow the DWIA calculation used in Refs.~\cite{Wakasa17,Yoshida22_Po}, only essential parts are described below.
The triple differential cross section (TDX) is given by
\begin{align}
  \frac{d^3\sigma}{dE_1^\mathrm{A} d\Omega_1^\mathrm{A} d\Omega_\mathrm{C}^\mathrm{A}}
  &=
  F_\mathrm{kin}^\mathrm{A}
  \frac{E_1 E_\mathrm{C} E_\mathrm{B}}{E_1^\mathrm{A} E_\mathrm{C}^\mathrm{A} E_\mathrm{B}^\mathrm{A}}
  \frac{(2\pi)^4}{\hbar v_\alpha}
  \frac{1}{2l+1}
  \frac{(2\pi\hbar^2)^2}{\mu_{p\mathrm{C}}^2}
  \frac{d\sigma_{p\mathrm{C}}}{d\Omega_{p\mathrm{C}}}
  \left|\bar{T}\right|^2.
  \label{eq:tdx}
\end{align}
The labels $0, 1$, and $\mathrm{B}$ denote the incoming proton, the emitted proton, and the residual nucleus, respectively, and $\mathrm{C}$ is the struck particle ($\mathrm{C} =$ N or $\alpha$). 
$E_i$ and $\Omega_i$ are the total energy and the solid angle of the emission direction of the outgoing particle $i = 1, \mathrm{B}$ or $\mathrm{C}$.
Quantities with a superscript $\mathrm{A}$ are evaluated in the nucleus rest frame (A frame), while the others are evaluated in the center-of-mass frame of the reaction.
$v_\alpha$, $l$, and $\mu_{p\mathrm{C}}$ are the relative velocity of the initial proton and the target, the orbital angular momentum of the $\mathrm{B}$-$\mathrm{C}$ bound state, and the reduced mass of the proton and $\mathrm{C}$, respectively.
The phase volume factor for this expression is denoted by $F_\mathrm{kin}^\mathrm{A}$; see Ref.~\cite{Yoshida22_Po} for the explicit expression.
$d\sigma_{p\mathrm{C}}/d\Omega_{p\mathrm{C}}$ is the differential cross section of the $p$-$\mathrm{C}$ elastic scattering; the energy and the scattering angle of the $p$-$\mathrm{C}$ binary scattering are determined from the ($p,p\mathrm{C}$) kinematics.

The reduced transition matrix $\bar{T}$ is given by
\begin{align}
  \bar{T} 
  &=
  \int 
  \chi_1^{*(-)}(\bm{R})
  \chi_2^{*(-)}(\bm{R})
  \varphi_{\mathrm{C}}(\bm{R})
  \chi_0^{ (+)}(\bm{R})
  e^{-\bm{K}_0\cdot\bm{R}A_\mathrm{C}/A}
  d\bm{R},
  \label{eq:tbar}
\end{align}
where $\chi_0$, $\chi_1$, and $\chi_2$ are the distorted waves between $p$-A, $p$-B, and C-B, respectively.
The superscripts $(+)$ and $(-)$ denote the outgoing and incoming boundary conditions, respectively.
$\varphi_{\mathrm{C}}$ is either the nucleon s.p. wave function ($\mathrm{C} =$ N) or the relative wave function between the $\alpha$ cluster and the residual nucleus B ($\mathrm{C} = \alpha$).
$\bm{K}_0$ is the momentum (wave number) of the incident proton. 
The mass numbers of particle $\mathrm{C}$ and the target nucleus A are given by $A_\mathrm{C}$ and $A$, respectively.
The total knockout cross sections discussed below are obtained by integrating Eq.~(\ref{eq:tdx}) over all possible kinematics.
In PWIA calculations, all the distorted waves in Eq.~(\ref{eq:tbar}) are replaced with the corresponding plane waves.

The DW-PW ratio is defined by
\begin{align}
  \frac{\sigma^\mathrm{DW}}{\sigma^{\mathrm{PW} }}
  &=
  \frac{ \displaystyle\int\left( 
    \displaystyle\frac{d^3\sigma}{dE^\mathrm{A}_1 d\Omega^\mathrm{A}_1 d\Omega^\mathrm{A}_\mathrm{C}}\right)
    dE^\mathrm{A}_1 d\Omega^\mathrm{A}_1 d\Omega^\mathrm{A}_\mathrm{C} 
  }{\displaystyle\int\left( 
    \displaystyle\frac{d^3\sigma^\mathrm{PW}}{dE^\mathrm{A}_1 d\Omega^\mathrm{A}_1 d\Omega^\mathrm{A}_\mathrm{C}}\right)
    dE^\mathrm{A}_1 d\Omega^\mathrm{A}_1 d\Omega^\mathrm{A}_\mathrm{C} },
  \label{eq:dwpwratio}
\end{align}
where $d^3\sigma^\mathrm{PW}/dE^\mathrm{A}_1 d\Omega^\mathrm{A}_1 d\Omega^\mathrm{A}_\mathrm{C}$ is the same as the TDX given by Eq.~(\ref{eq:tdx}) but calculated with PWIA.

\subsection{Numerical input}
For ($p,p$N) calculations, proton beams at 50~MeV and 200~MeV are considered.
The nucleon s.p. wave function $\varphi_N$ is obtained using the Woods-Saxon shaped potential proposed by Bohr and Mottelson~\cite{Bohr69}.
The depth of the potential is adjusted to reproduce the nucleon separation energy.
The global optical potential for a wide range of the proton nucleus scattering by Koning and Delaroche~\cite{Koning03} is employed to calculate the nucleon-nucleus distorted waves.
The $p$-N differential cross section is calculated using the Franey-Love effective NN interaction~\cite{Franey85}.
For the target nuclei, we consider $^{12}$C, $^{40}$Ca, $^{58}$Ni, $^{76}$Se, $^{90}$Zr, $^{108}$Cd, $^{120}$Sn, $^{144}$Er, $^{186}$Os, $^{196}$Pt, and $^{208}$Pb. 
We calculate the DW-PW ratio of Eq.~\eqref{eq:dwpwratio} for the 0p, 0d, 0f, 0g, 1p, 2s, and 2p nucleon s.p. orbitals, corresponding to ($n$,$l$) = (0,1), (0,2), (0,3), (0,4), (1,1), (2,0), and (2,1), respectively.

For the ($p,p\alpha$) calculations, following the former experiment~\cite{Carey84}, ($p, p\alpha$) reactions at 101.5~MeV on the target nuclei $^{12}$C, $^{16}$O, $^{20}$Ne, $^{24}$Mg, $^{28}$Si, $^{32}$S, $^{40}$Ca, $^{48}$Ti, $^{54}$Fe, $^{66}$Zn, $^{90}$Zr, $^{120}$Sn, and $^{208}$Pb are considered.
As mentioned in the previous subsection, the framework for the ($p,p\alpha$) reaction is basically the same as the ($p,p$N) reaction, except that the s.p. wave function $\varphi_\mathrm{N}$ is replaced with $\varphi_\mathrm{\alpha}$.
$\varphi_\alpha$ is obtained as a bound-state wave function of the Woods-Saxon potential.
Its range parameter $r_0 = 1.25B^{1/3}$~fm and the diffuseness parameter $a_0 = 0.76$~fm are taken from Ref.~\cite{Fukui16}.
Here, $B = A-4$ is the mass number of the core (residual) nucleus.
The optical potential parameters proposed by Avrigeanu and his collaborators~\cite{Avrigeanu94} are used for the distorted wave of the emitted $\alpha$.
The $p$-$\alpha$ differential cross section in free space is calculated by the folding model~\cite{Toyokawa13} using the Melbourne $g$-matrix interaction~\cite{Amos00} and a phenomenological $\alpha$ density.
For $\varphi_\alpha$, $l = 0$ is assumed, and the radial quantum number (the number of nodes) $n$ is determined such that the $\mathrm{B}+\alpha$ bound state has the same total harmonic oscillator quantum numbers of nucleus A in the ground state.
The specific values of $n$ are listed in Table~\ref{tbl:n_alpha}.

\begin{table}[!htb]
  \footnotesize
  \caption{The target nucleus and the radial quantum number $n$ of $\varphi_\alpha$ for the ($p,p\alpha$) knockout reaction.}
  \begin{tabular}{c||cc|ccccc|ccc|cc|c}
  \hline
  Nucleus & \multicolumn{1}{c|}{$^{12}$C} & $^{16}$O & \multicolumn{1}{c|}{$^{20}$Ne} & \multicolumn{1}{c|}{$^{24}$Mg} & \multicolumn{1}{c|}{$^{28}$Si} & \multicolumn{1}{c|}{$^{32}$S} & $^{40}$Ca & \multicolumn{1}{c|}{$^{48}$Ti} & \multicolumn{1}{c|}{$^{54}$Fe} & $^{66}$Zn & \multicolumn{1}{c|}{$^{90}$Zr} & $^{120}$Sn & $^{208}$Pb \\ \hline
  $n$   & \multicolumn{2}{c|}{2}                   & \multicolumn{5}{c|}{4}                                                                                                                       & \multicolumn{3}{c|}{6}                                                      & \multicolumn{2}{c|}{8}                      & 11         \\ \hline
  \end{tabular}
  \label{tbl:n_alpha}
\end{table}

\section{Results and discussion}
\subsection{Absorption effect and DW-PW ratio}
\begin{figure}[htp]
  \begin{center}
  \includegraphics[width=12cm]{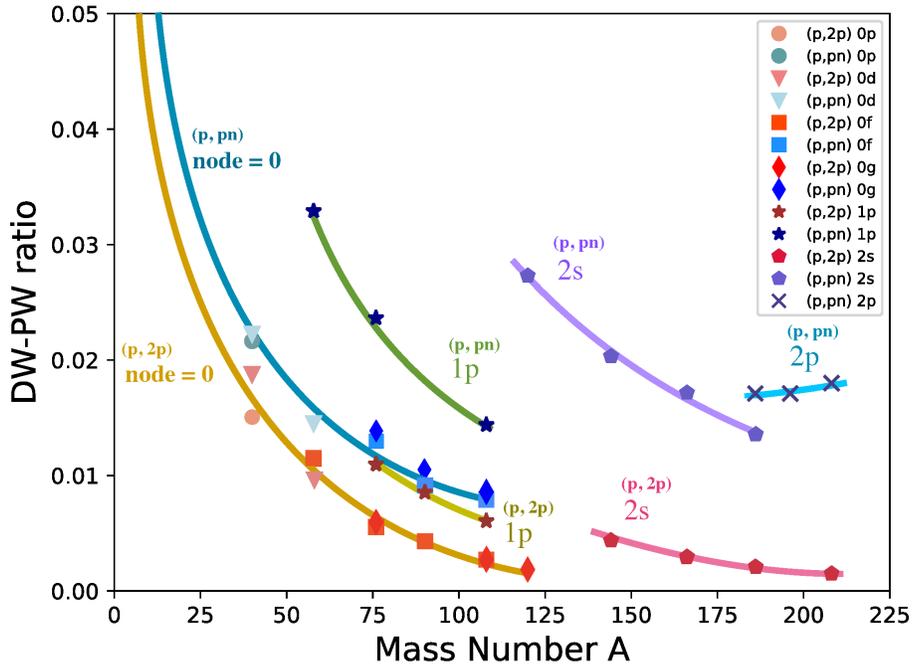}
  \end{center}
  \caption{DW-PW ratio of ($p$,$p$N) at 50 MeV.}
\label{fig:1}
\end{figure}

\begin{figure}[htp]
  \begin{center}
  \includegraphics[width=12cm]{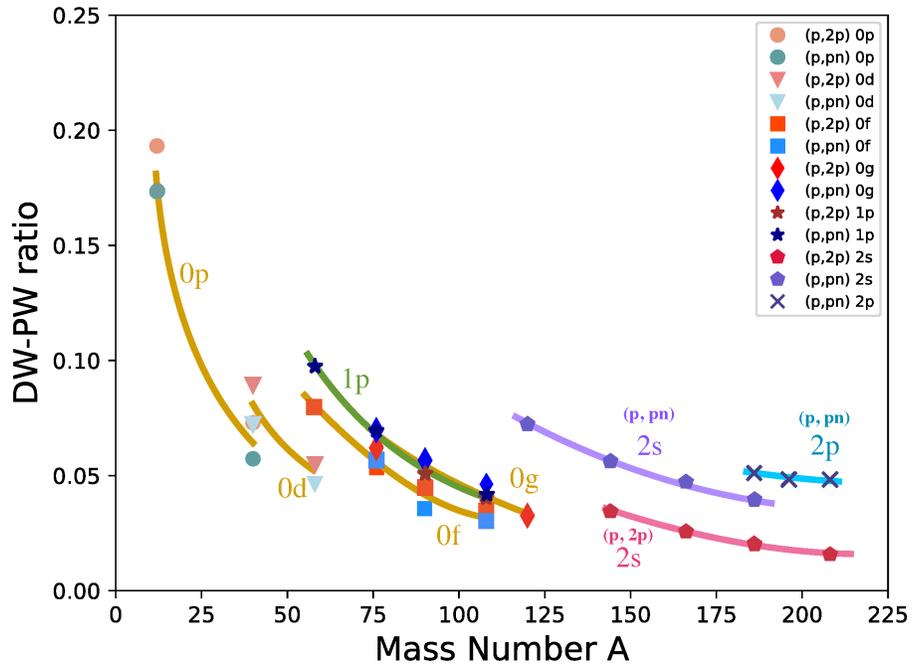}
  \end{center}
  \caption{DW-PW ratio of ($p$,$p$N) at 200 MeV.}
\label{fig:2}
\end{figure}

\begin{figure}[htp]
  \begin{center}
  \includegraphics[width=12cm]{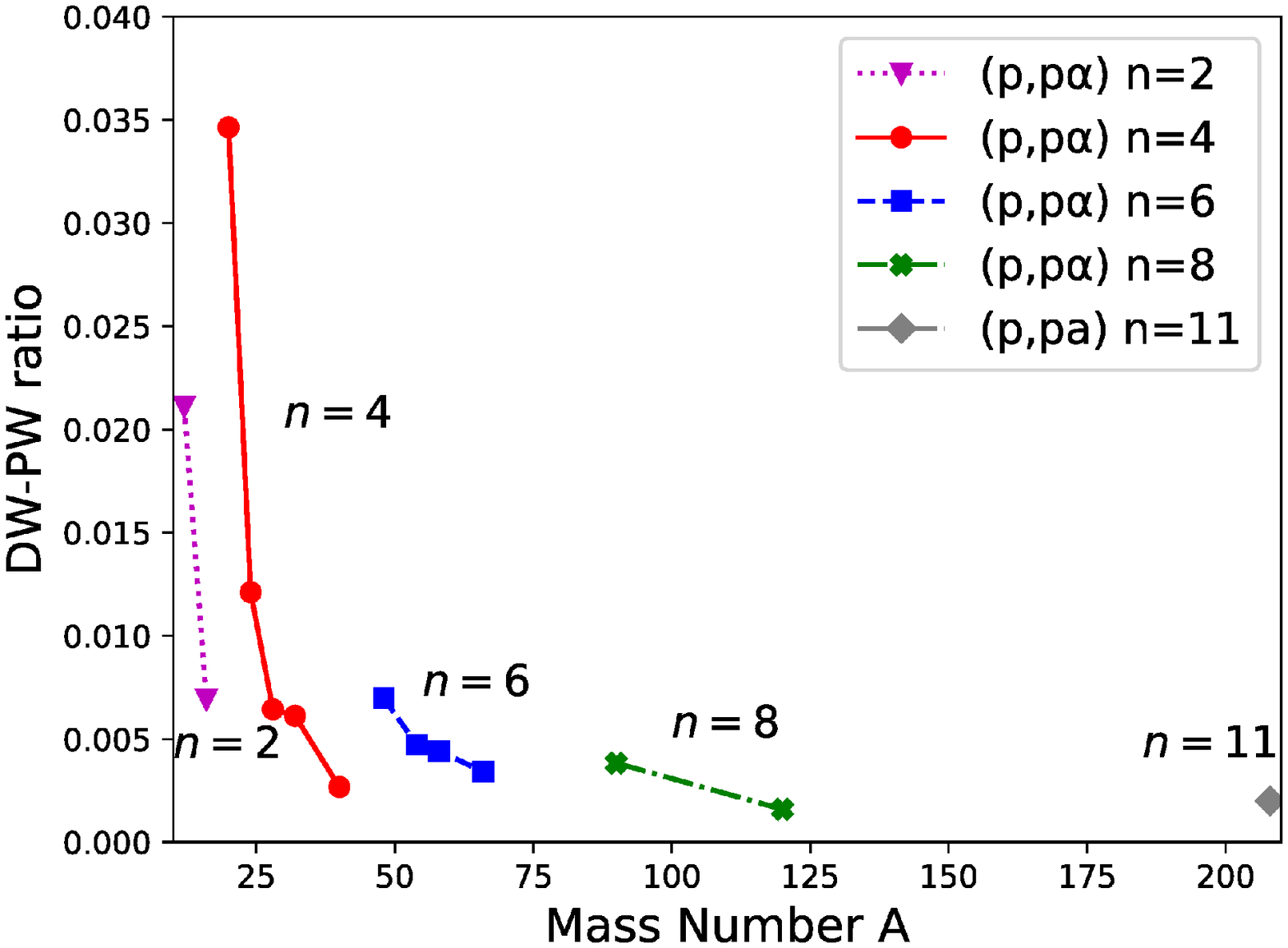}
  \end{center}
  \caption{DW-PW ratio of ($p$,$p \alpha$) at 101.5 MeV.}
\label{fig:3}
\end{figure}

The DW-PW ratio of the ($p,p$N) cross sections at 50 and 200 MeV are shown in Figs.~\ref{fig:1} and \ref{fig:2}, respectively.
It is found that the heavier the atomic nucleus, the smaller the DW-PW ratio, which is consistent with the result of the aforementioned Fig.~31 of Ref.~\cite{Wakasa17}. 
It can be also seen that the DW-PW ratio becomes larger as $n$ and $l$ increase, and particularly, it is more sensitive to changes in $n$.
The DW-PW ratio of the ($p, p\alpha$) cross sections at 101.5 MeV is shown in Fig.~\ref{fig:3}. Its dependence on the mass number $A$ and $n$ is similar to that for ($p,p$N).
These features of the DW-PW ratio can be understood as follows.

For given $n$ and $l$, the absorption effect becomes larger as $A$ increases. This is because the nonelastic processes for the scattering particles, i.e., particles 0, 1, and C, become more important for heavier target nuclei, reflecting the increase in the degree of freedom of the reaction systems. This general property of the DW-PW ratio can be seen in Figs.~\ref{fig:1}, \ref{fig:2}, and \ref{fig:3}. The only exception is the 2p orbital in the ($p$,$pn$) knockout reaction at 50 MeV; we will return to this point below.
The dependence of the DW-PW ratio for ($p,p$N) on the incident energy $E_{\rm in}$ can be understood in the same way as above. Namely, particles 0, 1, and C can easily penetrate the nucleus at higher $E_{\rm in}$. The absolute values of the DW-PW ratio in Fig.~\ref{fig:2} are noticeably larger than in Fig.~\ref{fig:1}.

As for the s.p. orbital dependence, the increase in the DW-PW ratio for higher $n$ is confirmed in the ($p$,$p$N) reactions at 50 MeV and 200 MeV and ($p$,$p\alpha$) at 101.5 MeV. This can be understood from the form of $\varphi_N$ and $\varphi_\alpha$. Their probability densities increase near the surface of the nucleus as $n$ increases. For similar reasons, the DW-PW ratio becomes larger when $l$ increases, but the change is not as significant as that due to an increase in $n$.
For $n=0$, the $l$-dependence of the DW-PW ratio at 50 MeV is found to be less noticeable compared to at 200 MeV.

For intuitive understanding of these features of the DW-PW ratio, let us introduce two radii: the absorption radius $r_{\rm abs}$ and the radius $r_{\rm C}$ of the nucleon or alpha particle to be knocked out. 
Here, we assume $r_{\rm C}$ is same as the root mean square radius of $\varphi_N$ and $\varphi_\alpha$.
Then, one may use the following geometrical model for the DW-PW ratio:
\begin{align}
  \frac{\sigma^\mathrm{DW}}{\sigma^{\mathrm{PW} }}
  &\sim
  \frac{r^3_{\rm C} - r^3_{\rm abs}}{r^3_{\rm C}}
  =1- \frac{r^3_{\rm abs}}{r^3_{\rm C}}.
  \label{eq:ratiomodel}
\end{align}
When we consider specific $n$ and $l$, $r_{\rm C}$ can be assumed to be a constant because its $A$-dependence is usually weak within a certain rage of $A$. 
As $A$ becomes larger or $E_{\rm in}$ becomes lower, $r_{\rm abs}$ increases, resulting in a smaller DW-PW ratio. 
On the other hand, for a specific $A$ and $E_{\rm in}$, $r_{\rm abs}$ can be regarded as a constant. 
Because $r_{\rm C}$ becomes larger as $n$ or $l$ increases, one obtains a larger DW-PW ratio. Now, let us consider the exceptional behavior of the DW-PW ratio for 2p neutron knockout at 50 MeV. In Table~\ref{tbl:radius}, we list $r_{\rm C}$ for 2p neutron considered in Fig.~\ref{fig:1} as well as for 2s neutron and 2s proton in the same mass region.
\begin{table}[!htb]
  \caption{The knocked-out nucleon radius $r_{\rm C}$ for the $A>150$ nuclei.}
  \begin{tabular}{c||ccc}
  \hline
  \multirow{2}{*}{Nucleus} & \multicolumn{3}{c}{$r_{\mathrm{C} }$}                                 \\ \cline{2-4} 
                           & \multicolumn{1}{c|}{2p neutron} & \multicolumn{1}{l|}{2s neutron} & 2s proton \\ \hline
  $^{166}$Er               & \multicolumn{1}{c|}{$\cdot$}    & \multicolumn{1}{c|}{5.77 fm}    & 5.25 fm  \\ 
  $^{186}$Os               & \multicolumn{1}{c|}{6.16 fm}    & \multicolumn{1}{c|}{5.86 fm}    & 5.35 fm   \\ 
  $^{196}$Pt               & \multicolumn{1}{c|}{6.25 fm}    & \multicolumn{1}{c|}{$\cdot$}    & $\cdot$   \\
  $^{208}$Pb               & \multicolumn{1}{c|}{6.37 fm}    & \multicolumn{1}{c|}{$\cdot$}    & 5.42 fm   \\ \hline
  \end{tabular}
  \label{tbl:radius}
\end{table}
As seen, while the $A$-dependence of $r_{\rm C}$ for the 2s proton and 2s neutron is weak, that for the 2p neutron is exceptionally strong. 
This makes the behavior of the DW-PW ratio rather nontrivial, meaning that we need to consider the $A$-dependence of not only $r_{\rm abs}$ but also $r_{\rm C}$. 
The results shown in Figs.~\ref{fig:1} and \ref{fig:2} suggest that the $A$-dependence of $r_{\rm C}$ is more (less) important than of $r_{\rm abs}$ for the 2p neutron DW-PW ratio at 50 MeV (200 MeV).


\subsection{The separation of the ($p$,$2p$) and ($p$,$pn$) ratios}
\begin{figure}[htp]
  \begin{center}
  \includegraphics[width=15cm]{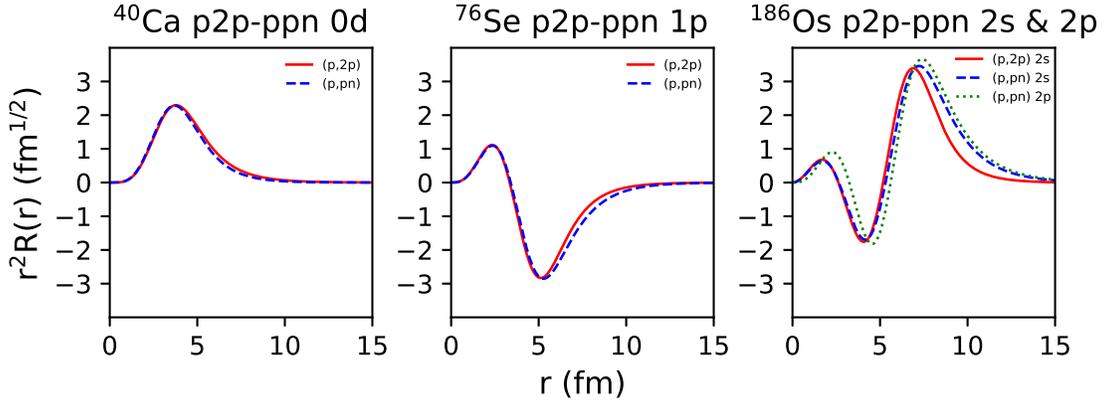}
  \end{center}
  \caption{Some selected s.p. wave functions $r^2 R(r)$ of $n$ = 0, 1, 2.
$R(r)$ is the radial part of the s.p. wave functions, and $r^2$ originates from $d\bm{R}$ in Eq.~\eqref{eq:tbar}. }
\label{fig:4}
\end{figure}

The difference between the DW-PW ratios of ($p$,2$p$) and ($p$,$pn$) appears in most s.p. orbitals at 50 MeV and in the 2s orbital at 200 MeV.
There are two main causes of this separation. 
One is the Coulomb potential for scattered nucleons, which makes it more difficult for protons to penetrate the nucleus at low energies, in particular, resulting in the reduction of the DW-PW ratio for ($p$,2$p$). 
At 50 MeV, this is the case with most orbitals, whereas, at 200 MeV, this effect is negligibly small.
The other cause of the separation of ($p$,2$p$) and ($p$,$pn$) is the s.p. wave function of the bound nucleons.  
In Fig.~\ref{fig:4}, some specific s.p. wave functions with $n$ = 0, 1, and 2 are shown. 
As the mass number increases, the difference in the s.p. wave functions of protons and neutrons develops because the Coulomb barrier suppresses the proton s.p. wave function at the nuclear surface.
This causes the separation of  the ($p$,2$p$) and ($p$,$pn$) for the 2s orbitals at 200 MeV.

The importance of these causes depends on the incident energy and $n$.
More specifically, the Coulomb potential provides the separation of $n$ = 0 orbitals at 50 MeV, 
and the difference in the bound-state wave functions gives the separation of the $n$ = 2 orbitals at 200 MeV. 
For the separation of $n$ = 1 and 2 orbitals at 50 MeV, these effects are combined.

\section{Summary and conclusions}
We conducted a systematic investigation of the absorption effect depending on the mass number and the orbitals of the bound proton, neutron, and $\alpha$ by calculating the ratio of the scattering cross sections of the DWIA and PWIA, i.e., the DW-PW ratio, for the ($p$,$p$N) and ($p$,$p\alpha$) reactions.
Since the PWIA cross section $\sigma^{\rm PW}$ is similar to the total elastic cross section of the elementary process, the DW-PW ratio tells us how much the knockout cross section is reduced from the cross section of the elementary process and will be helpful in estimating the experimental knockout cross section.
For the ($p$,$p$N) reaction, the target atomic nuclei $^{12}$C, $^{40}$Ca, $^{58}$Ni, $^{76}$Se, $^{90}$Zr, $^{108}$Cd, $^{120}$Sn, $^{144}$Er, $^{186}$Os, $^{196}$Pt, $^{208}$Pb, the s.p. orbitals of the bound nucleons ($n$,$l$) = (0,1), (0,2), (0,3), (0,4), (1,1), (2,0), (2,1), and the incident proton of 50 MeV and 200 MeV are considered.
For the ($p$,$p\alpha$) reaction, the target atomic nucleus $^{12}$C, $^{16}$O, $^{20}$Ne, $^{24}$Mg, $^{28}$Si, $^{32}$S, $^{40}$Ca, $^{48}$Ti, $^{54}$Fe, $^{66}$Zn, $^{90}$Zr, $^{120}$Sn, $^{208}$Pb, and the incident proton of 101.5 MeV are considered. 
In addition, the orbitals of the bound $\alpha$ with $l=0$ and $n = 2$, 4, 6, and 11 are assumed.
The mass number and s.p. orbital dependences of the absorption effect of the atomic nucleus were discussed by using the DW-PW ratio. 
As has been confirmed in the former review~\cite{Wakasa17}, in general, the heavier nuclei with higher mass numbers the smaller the DW-PW ratio in both ($p$,$p$N) and ($p$,$p\alpha$) reactions. 
This is because it becomes difficult for nucleons to penetrate the atomic nucleus, and the absorption effect of the atomic nucleus becomes stronger.
Regarding the dependence on the s.p. orbitals, the DW-PW ratio increases with increasing $n$ in both ($p$,$p$N) and ($p$,$p\alpha$) reactions, because the probability density of the s.p. state increases near the surface of the nucleus as $n$ increases, making it easier to scatter compared to penetrating the core of the nucleus. 
For similar reasons, the DW-PW ratio becomes larger with increasing $l$ in the ($p$,$p$N) reaction, but this change is less significant than its dependence on $n$ and is difficult to notice at around 50 MeV.
Finally, the difference between the DW-PW ratio of ($p$,2$p$) and ($p$,$pn$) reactions was discussed. 
At 50 MeV, clear differences in the DW-PW ratio between ($p$,2$p$) and ($p$,$pn$) reactions are observed in most s.p. orbitals. 
This separation is attributed to differences in the Coulomb potential for the scattered particle, that of the s.p. wave function, and the combination of these two.

Since the DW-PW ratio investigated in this study is for the s.p. total cross section, the contributions of each part of the atomic nucleus considering the angle and depth to the DW-PW ratio of knockout reactions were not fully understood. 
Further research on the DW-PW ratio for each part of the nucleus is needed for a better understanding of the absorption effect of the nucleus, such as the reversal of the $n=0$ and 1 orbitals at low incident energy~\cite{Wakasa17}, and the separations in the 0p and 0d orbitals at the incident energy of 200 MeV in the present work. 

\section*{Acknowledgments}
The authors are grateful to Y. Chazono (RIKEN) for valuable and helpful discussions.
The work of S.-I. Shim was supported by the National Research Foundation of Korea (NRF) grants funded by the Korean government (MSIT) (2018R1A5A1025563, 2022R1A2C1003964, and 2022K2A9A1A0609176).
This work has been supported in part by Grants-in-Aid of the Japan Society for the Promotion of Science (Grants No.~JP20K14475, No.~JP21H00125, and No.~JP21H04975).
\bibliography{ref}
\end{document}